\documentclass[12pt,journal]{IEEEtran}

\usepackage{epsfig}

\begin{document}


\title{ 
Data processing model for the CDF experiment }

\author{  \begin{center}
  J.~Antos,
  M.~Babik,
  D.~Benjamin,
  S.~Cabrera,
  A.W.~Chan,
  Y.C.~Chen,
  M.~Coca,
  B.~Cooper,
  S.~Farrington,
  K.~Genser,
  K.~Hatakeyama,
  S. Hou,
  T.L.~Hsieh,
  B.~Jayatilaka,
  S.Y.~Jun,
  A.V.~Kotwal,
  A.C.~Kraan,
  R.~Lysak,
  I.V.~Mandrichenko,
  P.~Murat,
  A.~Robson,
  P.~Savard,
  M.~Siket,
  B.~Stelzer,
  J.~Syu,
  P.K.~Teng,
  S.C.~Timm,
  T.~Tomura,
  E.~Vataga, and
  S.A.~Wolbers
\end{center}
\thanks{ J.~Antos, and M.~Babik are with 
         Institute of Experimental Physics, 
         Slovak Academy of Sciences, Slovak Republic.}
\thanks{ D.~Benjamin, M.~Coca, and A.V.~Kotwal are with 
         Duke University, Durham, NC 27708, USA. }
\thanks{ S.~Cabrera is with IFIC(CSIC-University of Valencia) E-46071 
         Valencia - Spain, visitor in CDF with Duke University. } 
\thanks{ A.W.~Chan, Y.C.~Chen, S.~Hou, T.L.~Hsieh, R.~Lysak, M.~Siket, 
         and P.K.~Teng are with Institute of Physics, 
         Academia Sinica, Nankang, Taipei, Taiwan.}
\thanks{ B.~Cooper is with University College London, 
         London WC1E 6BT, United Kingdom. }
\thanks{ S.~Farrington is with Liverpool University, 
         Liverpool L69 7ZE, United Kingdom. }
\thanks{ K.~Genser, I.V.~Mandrichenko, P.~Murat, J.~Syu, S.C.~Timm, 
         and S.A.~Wolbers are with
         Fermi National Accelerator Laboratory, Batavia, IL 60510 USA.}
\thanks{ K.~Hatakeyama is with The Rockefeller University, 
         New York, NY 10021, USA. }
\thanks{ B.~Jayatilaka is with University of Michigan, 
         Ann Arbor, MI 48109, USA. }
\thanks{ S.Y.~Jun is with Carnegie Mellon University, 
         Pittsburgh, PA 15213, USA. }
\thanks{ A.C.~Kraan is with University of Pennsylvania, 
         Philadelphia, PA 19104, USA. }
\thanks{ A.~Robson is with Glasgow University, 
         Glasgow G12 8QQ, United Kingdom. }
\thanks{ P.~Savard is with University of Toronto, 
         Toronto M5S 1A7, Canada.}
\thanks{ B.~Stelzer is with University of California, 
         Los Angeles, Los Angeles, CA 90024, USA. }
\thanks{ T.~Tomura is with University of Tsukuba, 
         Tsukuba, Ibaraki 305, Japan. }
\thanks{ E.~Vataga is with University of New Mexico, 
         Albuquerque, NM 87131, USA. }
}
\maketitle

\begin{abstract}

The data processing model for the CDF experiment is described.
Data processing reconstructs events from parallel data streams taken 
with different combinations of physics event triggers and further 
splits the events into datasets of specialized physics datasets.
The design of the processing control system faces strict requirements
on bookkeeping records, which trace the status of data files and 
event contents during processing and storage.
The computing architecture was updated 
to meet the mass data flow of the Run II data collection,
recently upgraded to a maximum rate of 40 MByte/sec.
The data processing facility consists of a large cluster of Linux 
computers with data movement managed by the CDF data handling system
to a multi-petaByte Enstore tape library.
The latest processing cycle has achieved a stable speed of 
35 MByte/sec (3 TByte/day).
It can be readily scaled by increasing CPU and data-handling capacity
as required.

\end{abstract}


\section{Introduction}

High-energy physics has advanced over the years through the use of 
higher energy and higher intensity particle beams and more capable 
detectors leading to the collection of larger volumes of data.
The Tevatron Run II project has increased the intensity and 
energy of the proton and anti-proton beams \cite{TeV2}.
The Collider Detector at Fermilab (CDF) experiment
has improved its data acquisition capacity in the Run II program
and is committed to studying the frontier of particle physics 
at Tevatron \cite{CDF2}.  
The computing facility was also upgraded for processing larger 
volumes of data.

In this paper we first describe the CDF data acquisition system with
parallel streams of raw data being logged to a mass storage library 
managed by the Enstore software system \cite{Enstore}. 
Section 2 describes the trigger system and the organization
of data streaming.
Data processing further splits the events into datasets of 
various physics interests.
The CDF computing requirement and the usage of
PC cluster are discussed in section 3.
The computing binary job is organized by collecting the required 
executable and software library in an archived file. 
It is suitable for submission in a distributed computing environment. 
The data processing architecture is described in section 4.

The CDF data files on tapes are registered to a Data File Catalog
(DFC) \cite{DFC}.  Data access has recently been migrated from direct 
tape read to the SAM (Sequential Access to data via Metadata) data 
handling system \cite{SAM}.
Prior to this change, data processing was operated with
the Farm Processing System (FPS) \cite{FPS}.
The PC cluster acting as the data production farm was upgraded with the 
CDF Analysis Farm (CAF) software \cite{CAF} which implements job 
submission and network access to the CDF data-handling system.
Details of the FPS and SAM production farm systems are described 
in Section 5.
The upgrade of the CDF data processing has incorporated the advances 
of SAM and CAF and is compatible with recent GRID computing development 
at Fermilab \cite{OSG-CAF}.

The design of the data processing includes task preparation, 
bookkeeping of input, output splitting and concatenation, and the
final storage to Enstore tape library.
The bookkeeping is required to ensure that there is no missing or 
duplicated data.
For the FPS farm, it is achieved using an internal database and
for the SAM farm by file metadata registered in the SAM database,
and is discussed in Section 6.
The computing resource management and recovery of production errors
are also discussed.
The processing capacity allows data to be reconstructed
in a timely manner.
The use cases described in Section 7 include detector monitoring, 
calibration, and physics data processing.
The experience of CDF data processing is discussed in Section 8.
The SAM farm processing is scalable by increasing parallel processes
to the network and the Enstore access limit.
Data processing capacity and scalability are discussed in 
Section 9.

\begin{figure}[bt!]
  \centering
  \epsfig{file=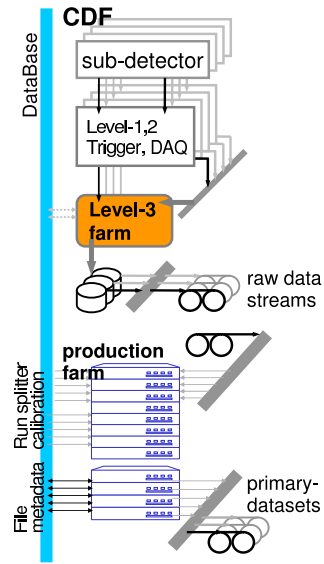,width=0.40\linewidth}
  \caption{CDF data of sub-detectors are pipelined 
   for selection of the level-1,2 trigger system.
   The complete event is reconstructed and selected by the Level-3 
   trigger.  Selected events are written to tapes in parallel streams.
   Data are retrieved and processed by the production farm. 
   The output is split and concatenated into 52 physics datasets.
  \label{fig:daq-tape} }
  \vspace{-.2cm}
\end{figure}

\begin{table}[b!]
  \begin{center}
  \begin{tabular}{|c|c|c|} \hline
   Stream & \multicolumn{2}{l|}{trigger contents } \\
   \cline{2-3}
     & event size (kByte) & ratio to total size (\%) \\
   \hline
   \hline
   A & \multicolumn{2}{l|}{monitoring}  \\
   \cline{2-3}
     &  195 &  2.5 \\
   \hline
   B & \multicolumn{2}{l|}{high $E_T$ leptons, missing $E_T$ } \\
   \cline{2-3}
     &  140 &  11.5 \\
   \hline
   C & \multicolumn{2}{l|}{high $E_T$ photons, di-photons } \\
   \cline{2-3}
     &  122 &  13.0 \\
   \hline
   D & \multicolumn{2}{l|}{study stream } \\
   \cline{2-3}
     &  344 &  11.2 \\
   \hline
   E & \multicolumn{2}{l|}{tau, lepton pair, Z to bb, higgs } \\
   \cline{2-3}
     &  146 &  16.0 \\
   \hline
   G & \multicolumn{2}{l|}{QCD jet, Di-jet, Miss $E_t$ plus jets, 
                          high $P_T$ b-jet } \\
   \cline{2-3}
     &  139 &  11.2 \\
   \hline
   H & \multicolumn{2}{l|}{$B$ to $\pi\pi$, $B_s$ to $D_s \pi$ } \\
   \cline{2-3}
     &  136 &  24.0 \\
   \hline
   J & \multicolumn{2}{l|}{$J/\psi$ to leptons, $\Upsilon$ to leptons}\\
   \cline{2-3}
     &  146 &  10.6 \\
   \hline
  \end{tabular}
  \caption{Statistics of data streams in 2005.
     Listed are the trigger contents, average event size and 
     the ratio to total size.
  \label{tab:rawdata} }
  \end{center}
  \vspace{-.5cm}
\end{table}

\section{Raw data recording }

The CDF detector is a large general purpose solenoidal detector
which combines precision charged particle tracking with fast projective
calorimetry and fine grained muon detection.
Measurements of track trajectories and energy depositions in
sub-detectors are collected for charged and neutral particles that 
result from the proton-anti-proton collision.
The data is organized into logical sets, each set being part of a 
``run''.
Each ``run'' corresponds to a time period during which the detector 
and beam conditions are stable.  
The readout electronics has a tiered trigger architecture
to accommodate a 396 ns proton-antiproton bunch crossing time.
To select only the most
interesting events a three level triggering system is used.
The data flow is illustrated in Fig.~\ref{fig:daq-tape}.
A pipelined Level-1 and Level-2 trigger is used before 
full events are selected for Level-3.
The Level-2 trigger accept flags an event for readout.
Data collected in DAQ buffers are then transferred to the Level-3
processor farm, where the complete event is assembled, analyzed, and,
if accepted, written out to permanent storage.
Output of Level-3 is handled by a Consumer-Server and Data-Logging 
subsystem which delivers events to mass storage and also to 
online consumer processes for monitoring purposes \cite{level3}.

The event triggers were chosen with the physics goals of the
experiment in mind.
Each event satisfies one or more triggers.  Events accepted 
by similar triggers are written to one of 8 data streams listed in 
Table~\ref{tab:rawdata}. 
Each of the 8 data streams will be further split into a total of 
52 datasets as part of the event processing stage.  
Each event is written to one or more of 
the 8 data streams based on the triggers that it satisfies.
The streams and datasets are chosen to maximize the physics utility 
of the datasets while at the same time reducing the number of events 
that are written to multiple data streams.  
The data is stored directly to Enstore tapes.

The raw data streams are for physics as well as for detector monitoring 
purposes.  The size of an physics event is on average 140 kByte.
Overlap between streams is about 5\% and increases with 
trigger rate.
The fractions of data streams vary: the largest one is for a dataset 
used to study physics related to charm and bottom quarks and it 
represents about 24\% of all accepted events.
Up to early 2006, a total of 2.4 billion events were taken
with total size of 387 TByte.  

\begin{figure}[b!]
  \vspace{-.2cm}
  \epsfig{file=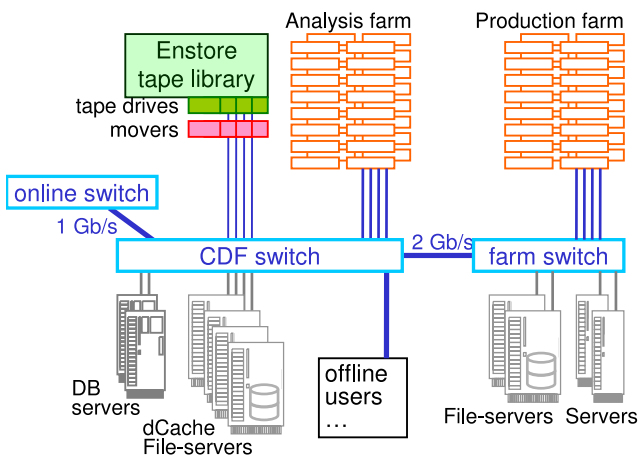,width=1.0\linewidth}
  \caption{The CDF production farm infrastructure (stale of 2006).
  \label{fig:farm-arch} }
\end{figure}

\section{Computing requirement}

The processing farm is required to reconstruct the
data collected by the experiment in a timely fashion.
The type of computing required for CDF data reconstruction can be 
characterized as loosely-coupled parallel processing~\cite{wolb}.
Each event is independent in the sense that it can be processed by
the offline reconstruction code without use of information from 
any other event.
The reconstruction program transforms digitized electronic signals 
from the CDF sub-detectors into information that can be used for 
physics analysis. 
The quantities calculated include particle trajectories and momentum, 
interaction vertex positions, energy deposition, 
and particle identities.

The CDF data processing facility is constructed using cost-effective 
dual CPU and multi-core CPU PCs running Linux.  The farm consists of 
a large number of PCs that run the CPU-intensive binary codes.
Data are buffered in and out of the farm using a disk cache system.
Data files are provided by the CDF data-handling and database server
for allocation and registration of files to the Enstore mass storage.
The hardware architecture of the CDF production facility is shown in 
Fig.~\ref{fig:farm-arch}.

The experiment requires rapid availability of data for data analysis 
with only a short delay to acquire the necessary detector 
calibrations.
To accomplish rapid data processing through the farms, adequate 
capacity in network and CPU is required.
From 2001 through 2004 the CDF experiment collected data 
at a peak throughput of 20 MByte/sec.
This was recently upgraded to 40 MByte/sec.
The event processing requires 2-5 CPU seconds on a Pentium III 1 GHz PC.
To accommodate the peak rate of 40 MByte/sec, corresponding to about 
25 million events (3.5 TByte) data logging a day, the total CPU 
required is about 1 THz.

\section{Data processing architecture}

The event reconstruction jobs are prepared using information from the
CDF database.
A binary job submitted to a worker node first receives
a binary ``tarball'', which is an archive of files created 
with the Unix tar utility.  It is self-contained, having the event 
reconstruction executable and all necessary libraries. 
The input file is copied from the CDF data-handling system to the
working area.  The required detector calibration and event trigger 
records are then accessed.
An input file is approximately 1 GByte in size and takes about 5 hours 
on a Pentium III 1 GHz machine to process.
The output is split into multiple files. Each file corresponds
to a primary dataset defined by the event type in the trigger system.
An event may satisfy several trigger patterns and is consequently 
written to multiple datasets that are consistent with that event's 
triggers.  The overlap of events in multiple datasets is about 6\%.

Once the processing job is successfully finished, all output files are 
copied to a cache area on a file-server where small files of the same
dataset are concatenated in run and event order to a 1 GByte 
file suitable for tape logging.
Depending on the event type, the size can vary from 20 kByte to over 
300 kByte. 
The output event has physics quantities added to the input, 
and is larger.
The total size of production output increases by 20\%.
For effective data handling, some of the datasets (of streams H and J)
are also split with compressed event contents, and the file size
is reduced by a factor of three.

The Run II CDF data-handling system has well-defined interfaces 
and operation \cite{data_handl} to provide input data for the farm and 
to write output to the Enstore mass storage system.
Raw data from the experiment is first written to tape in Enstore.
These tapes are registered in the Data File Catalog
as a set of tables in an Oracle database.
The data-handling system has been migrated to a SAM data-handling 
system with the file description recorded as metadata in a database.
The SAM data management is organized around a set of servers
communicating via CORBA \cite{corba} to store and retrieve files
and associated metadata.
Data files are registered and stored to required SAM cache and 
Enstore storage.
At Fermilab, the CDF data-handling system has a large dCache disk pool 
\cite{dCache} interface to Enstore storage.
By launching a SAM project with a predefined dataset,
SAM delivers files to the desired cache area.

\section{CDF production farm}

The CDF production farm performs computing and network intensive tasks 
in a cost-effective manner and is an early model for such computing.
Historically, Fermilab has used clusters of processors to provide large
computing power with dedicated processors (Motorola
68030)~\cite{gaines} or commercial UNIX workstations \cite{rinaldo}.
Commodity personal computers replaced UNIX workstations in the 
late 1990's.

The production farm development started in the late '90s.
The challenge in building and operating such a system is in 
managing the large flow of data through the computing units.
In addition, the production farm operation is required to be
easily manageable, fault-tolerant and scalable, with good monitoring and 
diagnostics.

CDF Run II data collected in 2001 to 2004 were processed by the first 
developed Farm Processing System (FPS) using the Fermilab developed 
FBSNG batch system \cite{FBS}. The disk cache used for input and
output files in process was a collection of ``dfarm'' file systems
\cite{dfarm} consisting space on the IDE disk drives on the workers.
An upgrade to the SAM data-handling system was conducted in 2004.
The SAM data-handling system is supported by Fermilab for file
cataloging and delivery in a distributed computing system.
The SAM production farm was streamlined with the FPS farm functions
packaged in a modular interface to the CAF computing facility
and SAM data-handling system.
The CAF is a Linux PC farm with interface modules for batch job
submission and access to the CDF data
management system and databases.
Jobs can be submitted to batch systems like FBSNG and Condor 
\cite{Condor} in a uniform manner.
The new SAM data production system is suitable for job submission to 
any computing facility in the world that uses CAF interface with 
direct access to the SAM data-handling and database connections. 
The production farm is thus a genuine Linux PC farm in a
shared computing environment with other CDF computing facilities
deployed at Fermilab and in many CDF collaboration institutes 
all over the world. Many of them are GRID computing facilities 
supplied with a CAF headnode and SAM stations.

\subsection{FPS farm} 

The Farm Processing System was the software that managed, controls 
and monitored the CDF production farm from 1999 to late 2005.  
It was designed to be flexible in configuration for production 
of datasets operated independently in parallel farmlets.
A farmlet is a subset of the farm resources specified 
for the input dataset, the executable and the output configuration 
for concatenation.
Its execution is handled by its own daemons taking care of 
consecutive processing in production and its records are written 
in the internal database.  

The FPS-farm had two server nodes hosting the FBSNG batch 
submission system, dfarm server, and the MySQL \cite{MySQL} server
used as the farm processing database.
The dual CPU workers were purchased over years 
with old nodes replaced after three years service.
At its peak in mid-2004, there were 192 nodes in service.
The dfarm used as the working cache was as large as 23 TByte 
including three file-servers each having 2 TByte.
The input and output (I/O) interface to the Enstore storage
was conducted by 16 Pentium nodes configured with the pnfs
file system \cite{pnfs} that accesses the Enstore tape library.
The number of workers and I/O nodes could be scaled 
to as large a number as was required and 
was chosen to match the total data through-put capacity 
and the number of Enstore movers (tape-drives) available.

\begin{figure}[t!]
  \centering
  \epsfig{file=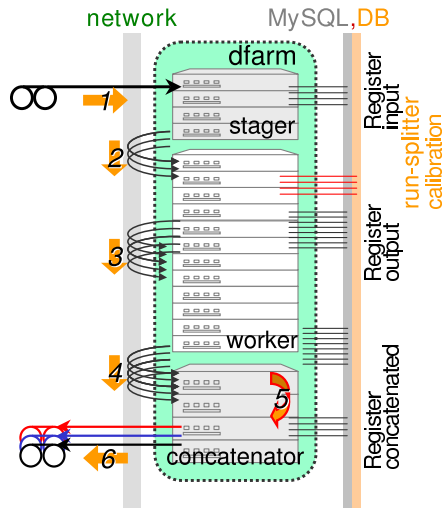,width=0.65\linewidth}
  \caption{FPS production farm architecture.
  \label{fig:fpsfarm} }
  \vspace{-.2cm}
\end{figure}

The production algorithm is logically a long pipeline with the 
constraint that files must be handled in order.
The data flow in FPS is illustrated in Fig.~\ref{fig:fpsfarm}.
The FPS tasks were conducted with control daemons running on the 
servers for resource management and job submission.
Monitoring and control interfaces for farm operation included a 
java server to the control daemons and a web server for monitoring.
The operation daemons were configured specifically for 
production of a input ``dataset'', usually of a specific
data taking periods of CDF.

The FPS production began with a farmlet menu selecting
input dataset and run range, tarball version, and output directories.
Input files were fetched directly from 
Enstore tapes and the outputs were written to output tapes.
The files moving through dfarm were controlled by four
production daemons.  
The daemons communicated with the resource manager daemon and 
the internal MySQL database to schedule job submission.
The internal database was used for task control and file-tracking. 
The tasks in sequence controlled by the daemons are :
\begin{itemize}
   \item {\bf Stager} is responsible for finding and 
   delivering files from tapes based on the farmlet menu 
   for data files in a run range in the dataset. 
   Jobs are typically submitted one ``file-set'' at a time.
   A file-set is a collection of files with a typical size of 10 GByte.
   The stager fetches DFC records for input and checks that
   proper calibration constants are available.
   The staging jobs are submitted to the input I/O nodes 
   and the file-sets are copied to their scratch area, 
   and afterwards to dfarm.

   \item {\bf Dispatcher} does the job submission through 
   batch manager to worker nodes.
   It looks for the staged input file, which is then
   copied into the worker scratch area.
   The binary tarball and run-splitter of trigger parameters are 
   also copied. 
   The reconstruction program runs locally on the 
   worker nodes and the output files are written locally.
   At the end of the job the output files are then copied back to dfarm.

   \item {\bf Collector} gathers histogram files, log files and any 
   additional relevant files to a place where members of the 
   collaboration can easily access them for the need of validation or 
   monitoring purposes.

   \item {\bf Concatenator} is responsible for collecting output
   files in run and event order to be concatenated into larger files 
   with a target file size of 1 GByte.
   It performs a similar task to the dispatcher
   according to the internal database records for a list of files
   and the concatenation jobs are submitted to output nodes.
   The output nodes collect files corresponding to a file-set size
   ($\approx 10$ GByte) from dfarm to the local scratch area and
   executes a merging program to read events in the input files
   in increasing order of run numbers.
   It has a single output truncated into 1 GByte files.
   These files are directly copied to tapes,
   and DFC records are written.
\end{itemize}

Since all of the farmlets shared the same sets of processors and
data storage, the resource management was
a vital function of FPS for distribution and prioritization
of CPU and dfarm space among the farmlets. 
The additional daemons are:
\begin{itemize}
   \item {\bf Resource manager} controls and grants allocations for 
   network transfers, disk allocations, CPU and tape access based on a 
   sharing algorithm that grants resources to each individual farmlet 
   and shares resources based on priorities.  This management of 
   resources is needed in order to prevent congestion either on the 
   network or on the computers themselves and to use certain resources
   more effectively.
        
   \item {\bf Dfarm inventory manager} controls usage of the 
   distributed disk cache on the worker nodes that serves as a 
   front-end cache between the tape pool and the farm.  

   \item {\bf Fstatus} is a daemon that checks periodically whether 
   all of the services that are needed for the proper functioning of 
   the CDF production farm are available and to check the status 
   of each computer in the farm. 
   Errors are recognized by this daemon and are reported 
   either to the internal database which can be viewed on the web or 
   through the user interfaces.
\end{itemize}

The FPS system status was shown in real time on a web page giving
the status of data processing, flow of data, and other useful 
information about the farm and data processing.
The web interface was coded in the PHP language \cite{php} and 
RRDtool \cite{rrd} for efficient storage and display of time 
series plots.
The structural elements in the schema include output from FPS modules, 
a parser layer to transform data into a format suitable for RRDtool,
a RRDtool cache to store this data in a compact way, and finally 
the web access to RRD files and queries from MySQL for real time 
display of file-tracking information.

The FPS framework ran on one of the servers and depended on the kernel 
services namely the FBSNG batch system, and the FIPC (Farm Interprocess 
communication) between the daemons and dfarm server governing 
available disk space on the worker nodes.  
Daemons had many interfacing components that allowed them to communicate
with the other needed parts of the offline architecture of the CDF 
experiment. 
Those included mainly the DFC (Data File Catalog) and the Calibration 
Database.

\begin{figure}[t!]
  \centering
  \epsfig{file=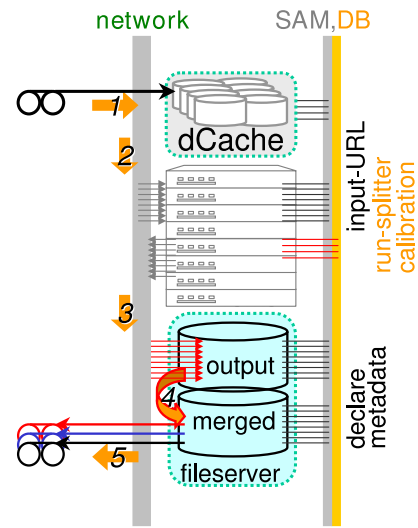,width=0.60\linewidth}
  \caption{SAM production farm architecture.
  \label{fig:samfarmflow} }
  \vspace{-.2cm}
\end{figure}

\subsection{SAM production farm }

A production farm upgrade was necessary to accommodate the 
increasing data acquisition rate and the migration of the CDF 
data-handling system to SAM.
The production farm was converted to a CAF facility with
direct access to the CDF dCache disk pool and Enstore tape storage.
The functions of the FPS farm were replaced by new modular applications
for job submission to CAF and file delivery performed by SAM 
applications.
The worker nodes are only used for job execution.
The dfarm file system was not used; instead, staging of input files
is conducted by SAM on a dedicated dCache disk pool.
The job outputs are sent to a ``durable storage'' on which the files 
reside briefly before being stored elsewhere.
The durable storage in use are disk RAIDs on file-servers equipped
with one or more giga-bit ethernet links.
In consideration of network bandwidth and disk cache usage,
the SAM production farm is configured for direct file copy from the 
dCache system where input files are staged from Enstore.
Concatenated output files are transferred directly to Enstore.

\begin{figure}[t!]
  \centering\epsfig{file=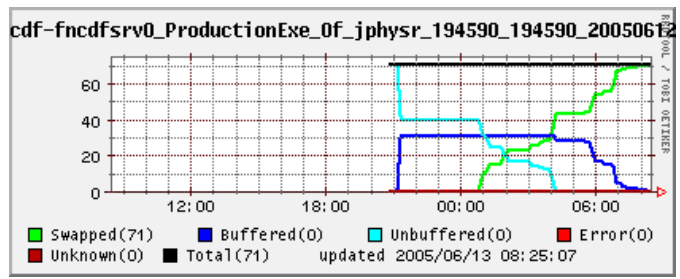,width=1.0\linewidth}
  \caption{The consumption of files by a SAM project is plotted
    for a dataset of 71 input files.
    Files were quickly "buffered" to CAF and the batch job was
    configured to use 30 CPU's.
    The executable lasted for approximately 4 hours. Consumed files 
    were being "swapped".  The project was terminated after all files 
    were swapped.
  \label{fig:project} }
\end{figure}

The production job submission is controlled by scheduled cron jobs 
on a server node that is configured as a SAM station to
issue SAM client tasks.
The SAM farm design is modular and allows for more flexible use of
worker nodes and storage.
The data flow in the SAM production farm is illustrated in 
Fig.~\ref{fig:samfarmflow}.
A production task is launched as a SAM project associated with a 
predefined user dataset containing the files to be processed.
\begin{itemize}
\item  {\bf Preparation of input datasets: } \\
       Input data to be processed are selected by querying raw data
       DFC records, readiness in data-handling system,
       and the presence of detector calibration.
       The input datasets are organized in run sequence 
       of one or multiple runs from a single raw data stream.

       The dataset bookkeeping list is updated continuously.
       The expected output files are checked for.
       A dataset not yet having complete output will be submitted
       to CAF.

\item  {\bf Starting a SAM project for CAF submission: } \\
       A user defined dataset not in use is checked to see whether 
       its output is complete. If not, a SAM project is launched and 
       a SAM consumer process is established for delivery of 
       files in the dataset to the dCache pool.
       A CAF submission follows for the associated project.
       The binary tarball is distributed to worker nodes in CAF.
       It is extracted by each process on the local working area
       and the task to be executed is given by a shell script.
       A SAM query is then made asking for the location of an input file
       on dCache. The input file is copied to the local working area. 
       The detector calibration and online split table are fetched. 
       The binary job is executed for one input data file.
       After the program is successfully terminated, all output 
       files are declared to SAM and are copied to the dedicated 
       durable storage nodes.

       The worker script asks for another file not yet consumed. 
       The same execution is practised until all files in dataset
       are consumed.  Afterwards, the project is closed,
       the CAF job terminates and all log files are copied to 
       servers.

       The SAM metadata of output files are used extensively for 
       bookkeeping purpose.  It has a parentage record of the input 
       file, and after concatenation will be updated with the file
       it is merged into.  The completeness of a dataset 
       is checked by querying all its output metadata.
  
\end{itemize}
Monitoring of a SAM project is illustrated in Fig.~\ref{fig:project},
which shows the file consumption status in time.
In this case input files are already buffered on dCache and are 
quickly distributed to workers on a CAF.  
The processing of an input file takes about 4 hours and the 
workers stay busy until all file are consumed.

The output of CAF jobs is sent to durable storage on file servers.
The durable storage consists of a total of 52 directories;
one for each primary physics dataset.
The concatenation job is controlled by a cron job which does a periodic 
check on the total number of files accumulated in a primary physics 
dataset.  If the number of files passes a threshold 
(for example, 100 files) a concatenation job is launched 
to merge files into output files of size close to 1 GByte.  
The operations in sequence are:
\begin{itemize}
\item  {\bf Preparation of file list:} \\
       The SAM metadata of files to be merged are checked to
       ensure none was processed previously.
       These files are sorted by run number into lists of files 
       with a constraint that the size of each output file is close 
       to and not exceeding 1 GByte.

\item  {\bf Concatenation job:} \\
       A concatenation binary is executed in order for files
       on each lists.  It does the file opening, block-move to the 
       merged output, until all files in a list are finished.
       The merged output files are sent to the ``merged'' directory 
       ready to be stored to SAM.
       SAM metadata is declared for the
       merged file recording the parents of their input files.
       The metadata of files being merged are updated recording the
       file they have entered.

\item  {\bf SAM store:} \\
       Merged files are scanned periodically by a cron job to check 
       whether they exceed a threshold (for example 10 files).
       If a file has not been stored, 
       the SAM store command is issued to copy it to Enstore
       mass storage and its metadata are declared.
       The threshold size is tuned to optimize Enstore operations.

       The metadata of a stored file will later be updated with its tape
       volume and location in the pnfs file system.
       A merged file having complete metadata records for its tape
       location is then deleted from durable storage disks.

\end{itemize}

The concatenation is issued with no constraint imposed
on the SAM project status.  This allows for
modular operation separating usage of durable storage 
from CAF computing activities.
In order to have concatenation files organized in run order,
the CAF jobs are scheduled such that the output arrives in order.
The threshold on the number of files required for launching 
concatenation is used as a throttle to allow sufficient time 
to wait for slow CPU nodes whose output arrives late.
Files missing as a result of other failures will be recovered 
automatically by the job submission, and with sufficient buffer 
size in the durable storage, the file order is organized
by the concatenation process.
Previously in the FPS operation, concatenation waited for 
specific output files of submitted jobs.
As the executable may crash and hardware does fail,
it resulted in unexpected congestion in data flow waiting for
missing files.
This is relaxed in the new concatenation algorithm,
checking only files that have arrived at durable storage.

The usage of large file servers for durable storage 
has provided robust file management.
This, however, has led to concern over collecting hundreds of files 
from CAF workers through giga-bit ethernet links.
To achieve maximum file movement speed, concatenation jobs run
locally on the file servers.
The final storage copying merged files to Enstore tape typically 
runs at 20 MByte/sec. 
A file server with dual Pentium 2.6 GHz CPUs 
can move, concatenate and SAM store to tape about 1 TByte/day.

\section{Bookkeeping, resource management and recovery } 

The production farm processes hundreds of files at a time and
effective bookkeeping is essential.
The contents of bookkeeping are the history of the files as they
flow through production,
relation of input and output, and status of binary reconstruction
and hardware.
The MySQL database was a good choice for the FPS system, and was used 
since the production farm development.
The file status in FPS was traced by job daemons in three tables.
Usage of these tables are:
\begin{itemize}
\item  {\bf Stage-in table: } \\
  A cron job does a periodic check on the farmlet run range,
  makes a query to the DFC database to update the stage-in table for
  input files to be processed.
  The stager daemon is operated according to this table to copy
  files from Enstore, and afterwards move their records off the 
  stage-in table to a history table.
\item  {\bf Reconstruction table: } \\
  The dispatcher daemon checks on staged files and their history
  in stage-in table.
  Files not yet processed are registered in this table and are
  later submitted to FBSNG. The file consumption status
  are then updated by the worker script.  
  Once a file is successfully processed, its records are copied 
  to the corresponding history table.
  Its output files are registered into the concatenation table to be 
  picked up by the concatenation daemon.
\item  {\bf Concatenation table: } \\
  A concatenation job is organized with a query to MySQL for
  processed files of approximately 10 GByte. 
  The concatenation process split output into 1 GByte files.
  Therefore an input file often is split into two concatenated files.
  This leads to a difficulty in recovery, if required,
  for tracing its input.
\end{itemize}
For debugging purpose, the FPS bookkeeping database has recorded
details on file delivery time and binary crash records.
These records are valuable to diagnose data handling and 
binary problems.  Use of these records provided valuable guidelines
in the upgrade to the SAM production farm.

\begin{figure}[t!]
  \centering\epsfig{file=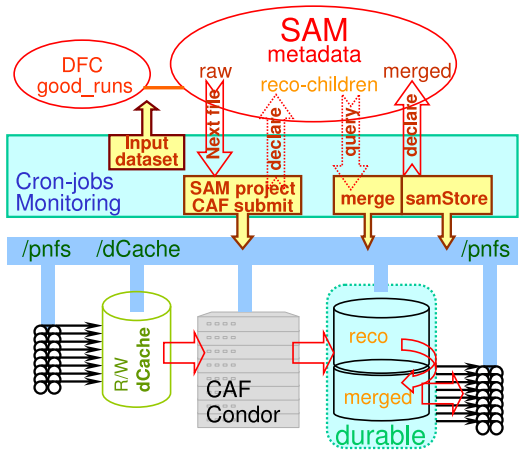,width=.8\linewidth}
  \vspace{-.4cm}
  \caption{Communication with online DFC records and SAM metadata 
    by the production tasks along the data flow from left to right.
  \label{fig:samfarm-arch2} }
\end{figure}

The bookkeeping for the SAM farm was streamlined
by appending file parentage records to metadata of files
in the SAM data handling system.
Job submission is based on the cumulative records of datasets in SAM 
and is therefore simpler than the FPS farm in tracking individual files.
The communication of production tasks with DFC records and 
SAM metadata is illustrated in Fig.~\ref{fig:samfarm-arch2}.
These operations are:
\begin{itemize}
\item  {\bf Input datasets: }\\
  Preparation of input datasets is conducted by a periodic cron job
  fetching online records for newly available data.
  New data are organized into SAM datasets that are retrieved by 
  the cron job doing CAF submission.
\item  {\bf Metadata of reconstruction output: }\\
  The completeness of an input dataset is checked for the expected 
  output.  Each output file has metadata created when it is successfully
  written.  Its records on parent and daughter files are the two tags 
  that help define its complete history in production. 
  The parent is the input file from which it is created,
  and the daughter is the merged file after concatenation.
\item  {\bf Metadata of concatenation output: }\\
  The reconstruction output files stored in durable storage
  are checked for the existence of their daughters in the
  concatenated files. Those not yet been merged are 
  collected for concatenation.  This procedure prevents
  duplication of production output.
  The metadata of a merged file inherits all the parents of its input
  from reconstruction.  This creates a complete bookkeeping chain
  for production datasets and outputs.
  The archive of parentage records summaries the production activity.
\end{itemize}
The bookkeeping of the SAM production farm has advanced from counting
individual files to datasets: each is submitted as a SAM project
to a corresponding CAF job.  
Therefore the monitoring of production is reduced 
to monitoring datasets in progress.
The resource management is thus focused on the number of 
live datasets, flow of data-handling, and the durable storage reserved.

Binary crashes and hardware failure always occur and are not easily
predictable.
To prevent congestion, in the SAM farm design a latch
mechanism was implemented by tracking the availability of resources.
The resource management falls into the categories of CPU, 
cache storage, database services and data-handling system.
The old FPS farm managed resources using
daemons that monitored their status.
To be robust, the SAM farm design is simplified with
a frequent cron job monitoring the resources and updating a status page.
New production tasks are prevented
if any of the required services and resources are missing.
The concatenation process is flexible and is independent of the CAF
usage.  
Jobs are not submitted if insufficient space is available on durable 
storage.

Having hundreds of files at a time moving through the system, it
is necessary to have automatic error recovery with
minimal human interaction in the course of processing.
This relies on the cron job scripts that check bookkeeping records 
for automated recovery and return to a state from which the farm 
can safely continue to operate.
The human effort is then spared for debugging abnormal failures
in hardware or other failures.
The parentage of metadata used for the SAM production farm provides
a convenient bookkeeping interface.
A concatenated file is a merge of full files.
It is easier to recover based upon its parentage list.

\section{Production farm use cases }

The mass data reconstruction is used for various purposes, including 
detector and data monitoring, calibration creation, processing for 
physics analysis, and reprocessing when that is required to take 
into account improvements in calibrations and/or software.
The three main categories of farm operation are:

\begin{itemize}
\item  {\bf Calibration:} \\
The most urgent calibration required by CDF is the determination 
of the proton-antiproton beam-line position.
This process is run as soon as new data is available in SAM.
The executable uses only a small portion of tracking data and output
histogram files are sent to a dedicated storage area for rapid
availability.

Meaningful physics data are processed with a complete set of detector 
calibrations that may require large statistics of certain types of 
events.
Calibration is a time consuming procedure because it requires careful
examination of the detector performance and understanding of variations
from what is expected.

\item  {\bf Detector monitoring:} \\
A monitoring data stream is processed instantly once the beam-line
calibration is available.  The purpose is for examination of detector 
performance.  Depending on the colliding beam quality and the presence 
of sub-detectors, the data quality is categorized.
Usage of physics datasets is defined accordingly.

\item  {\bf Physics data processing:} \\
Once the calibration data are available, the raw data streams are
processed.  For the convenience of bookkeeping and resource management, 
the data streams are scheduled in parallel.  In the SAM farm operation,
one raw data stream is processed by about 200 workers and the output 
from all of the workers is sent to one durable storage file server. 
The production farm throughput for each stream is about 1 TByte per day.

If required, the production farm also does reprocessing of primary 
datasets, to take advantage of improved binary code or calibrations.
In such case concatenation may not be required.

\end{itemize}

\section{Experience with the Production Farm}

The production farm has been in operation since the beginning of
CDF Run II data collection.
The Tevatron Run II commissioning run started in October, 2000 
and the beginning of proton-antiproton collisions in April, 2001. 
The data taken under various beam and detector conditions,
consist of about 7.6 million events.
The early processing experience using a prototype FPS production farm
gave some confidence that the farm had the capacity to handle 
the volume of data coming from the detector and also uncovered 
many operational problems that had to be solved.

Beginning in June, 2001, both the Tevatron and the CDF detector operated
well and began to provide significant data samples for offline 
reconstruction. This early data was written in 4 streams and 
the output of the farms was split into 7 output datasets. 
The CDF experiment wrote data at a peak rate of 20 MByte/sec, 
which met the design goal. The farms were able to reconstruct data at 
the same peak rate. The output systems of the farm were adjusted
to increase their capacity to handle the large output of the farms. 
More staging disk was added to provide a larger buffer and additional 
tape-drives were added.

By early 2002 the accelerator was running steadily and the CDF 
detector was performing well.  Data was recorded in 8 data streams 
and the output was split into 52 physics datasets.
Data was processed as quickly as possible with preliminary calibration
and was normally run through the farms within a few days.
Approximately 1.2 billion events were collected and processed 
between 2002 and 2004.
Upgrades were made to the farm with the addition 
of new nodes for processing as well as improved I/O capability.
A major reprocessing of all data collected from the beginning of 2002 
was begun in the fall of 2003.  The output of this processing was 
later reprocessed with improved calibration of calorimetry and 
tracking.
The reprocessing occurred in March, 2004 and the production farm 
was operated at full capacity at a rate of processing of 10 million 
events per day.

To accommodate the migration of CDF computing to a distributed
environment, the SAM farm upgrade was started in winter of 2004.
Building on the knowledge gained from the FPS farm performance, 
the upgrade advanced quickly. The SAM farm tests were completed 
in early 2005.  By the summer of 2005 the migration was 
completed.  The SAM farm processing, with its own group of 200 nodes
is as effective as the FPS farm and the operation is easier.

The SAM production farm was tested further on its scalability 
and on job submission to a distributed computing environment.
Test jobs were sent to remote CAF overseas to prove the mechanism.
For effective network access to dCache and Enstore storage.
The production jobs for 2005 data were conducted on two CAFs 
(SAM production farm and CDF analysis farm) at Fermilab.
The largest number of CPUs used was 560, and was not yet limited,
given a stable processing rate of over 20 million events a day.
A single CAF can be expanded to up to one thousand CPUs.
The bottleneck observed comes from the 1 Gbit/s network links writing to
durable storage.  This bottleneck can be easily eliminated by dividing 
the processing into more data processing streams and more file servers.

\section{Data processing capacity}  

\begin{figure}[t!]
  \centering\epsfig{file=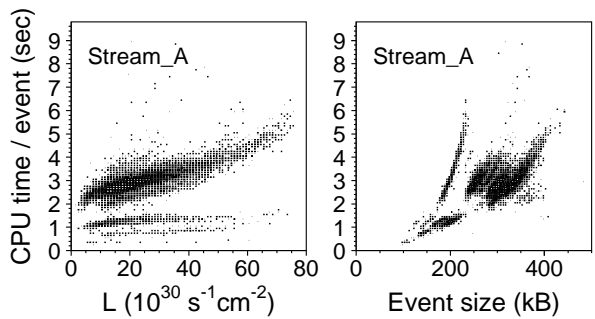,width=1.0\linewidth}
  \vspace{-1.cm}
  \caption{CPU time per event versus the proton-antiproton beam 
    luminosity and event size of the detector monitoring stream.
    The CPU time is normalized to Pentium III 1.0 GHz.
    The separation in bands is attributed to the presence of 
    the silicon detector (larger event size) and to event types.
  \label{fig:CPUevt} }
  \vspace{-.2cm}
\end{figure}

The CPU speed and data through-put rate are two factors that determine 
the data reconstruction capacity of the production farm.
The computing time required for an event depends on the event 
characteristics determined by the event trigger in different data 
streams and the intensity of the proton and antiproton beams.
More intense beams lead to multiple events per beam
crossing which in turn lead to more CPU time per event.  
The event size and CPU time varies for different raw data streams.
In Fig.~\ref{fig:CPUevt} the CPU time per event is illustrated
for the monitoring stream (Stream A) that has an average event size 
that is twice larger than the physics data streams.
The CPU time on a dual Pentium III 1 GHz machine varies
from 1 to 10 sec depending on the beam intensity and event size.

Inefficiency in utilizing CPU is primarily due to the execution time 
lost during file transfers of the executable and data files to and 
from the worker scratch area.
The data flow is ultimately limited by the capacity and speed 
to retrieve and store files in the Enstore tape storage.
Enstore files are read/written directly to/from the durable storage 
disk without a buffer,
the tape read/write has a latency of a few minutes to mount the tape 
in the tape driver.  For maximum efficiency many files
of the same dataset are written as a group.
The instantaneous Enstore read/write speed may exceed 30 MByte/sec.
However, the average rate drops to below 20 MByte/sec because of 
latency in establishing connection to the mass storage system (this
includes mounting and positioning the tape establishing the end-to-end
communication).
Depending on the latency spent on tape mounting and network traffic,
a steady tape writing rate by one mover for is about 1 TByte/day.

The output of concatenated files is copied to tapes.  
A tape is restricted to files of the same dataset.  
The tape writing is limited to one mover per dataset at a time, 
to ensure that files are written on tape sequentially.
The effectiveness in tape writing is a concern because of
the limited cache space and output bandwidth.
Concatenation by the FPS farm was conducted on output nodes.
On a Pentium 2.6 GHz node the CPU time is about 24 minutes for 
processing 10 GByte, followed by 10 minutes of copying concatenated 
files to tape.
In order to improve the effectiveness copying files of a dataset to 
tapes, up to three concatenation jobs for the same dataset were
submitted in series to eliminate the waiting time by the mover.
Also, running a mix of jobs from different datasets in parallel 
increases the through-put of the farm by increasing the number 
of movers writing tapes.

\begin{figure}[t!]
  \centering\epsfig{file=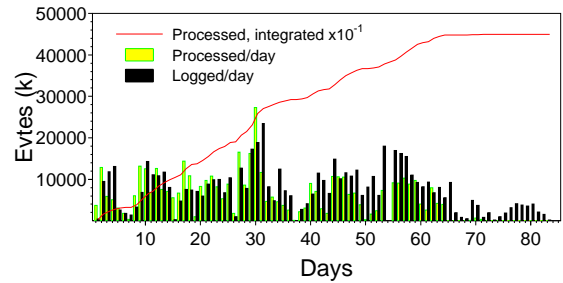,width=1.02\linewidth}
  \vspace{-1.cm}
  \caption{Daily processing and logging rates.
    The integrated processing rate is shown in line.
  \label{fig:stat531} }
  \vspace{-.2cm}
\end{figure}

The FPS farm capacity may be illustrated by the latest
data reprocessing of primary physics datasets performed in March 2004.
To maximize resource usage, the reprocessing was performed on five 
farmlets with each farmlet processing one dataset.
The stager was submitted one dataset at a time, therefore
the farm CPU usage came in waves of two or more datasets at a time.
The stage-in was effective in feeding data files to dfarm.
The CPU usage was efficient.
The data processing rate is shown in Fig.~\ref{fig:stat531}.
On average a through-put of 10 million events (1.5 TByte) 
per day to the Enstore storage was achieved.
The data logging lasted two extra weeks for a large B physics dataset
that accounted for about a quarter of the total CDF data. 
It was the latest dataset processed and was saturated at 
about 1 TByte per day in writing tapes.

The SAM production farm exploits the advantages of the new data 
handling system and the capacity of submitting jobs to multiple 
CAF farms.
The performance may be illustrated in data processing of 2005.
The production jobs were submitted to two CAFs with a total 
560 CPU processing 6 raw data streams in parallel.
The network through-put with workers of Pentium-4 3 GHz CPU
was roughly saturating at 200 CPU for one data stream with 
reconstruction output sent to durable storage on one file server
equipped with one 1 Gbit/s network port.
With more than 200 CPU engaged for one data stream,
the reconstruction output from workers was often waiting in queue 
to copy files to durable storage.
The file server equipped with 1 Gbit/s ethernet has a constant
traffic of over 50 MByte/sec shared by input and output.  
The data throughput rate on a file server was operated at
less than 1 TByte daily.

\begin{figure}[t!]
  \centering\epsfig{file=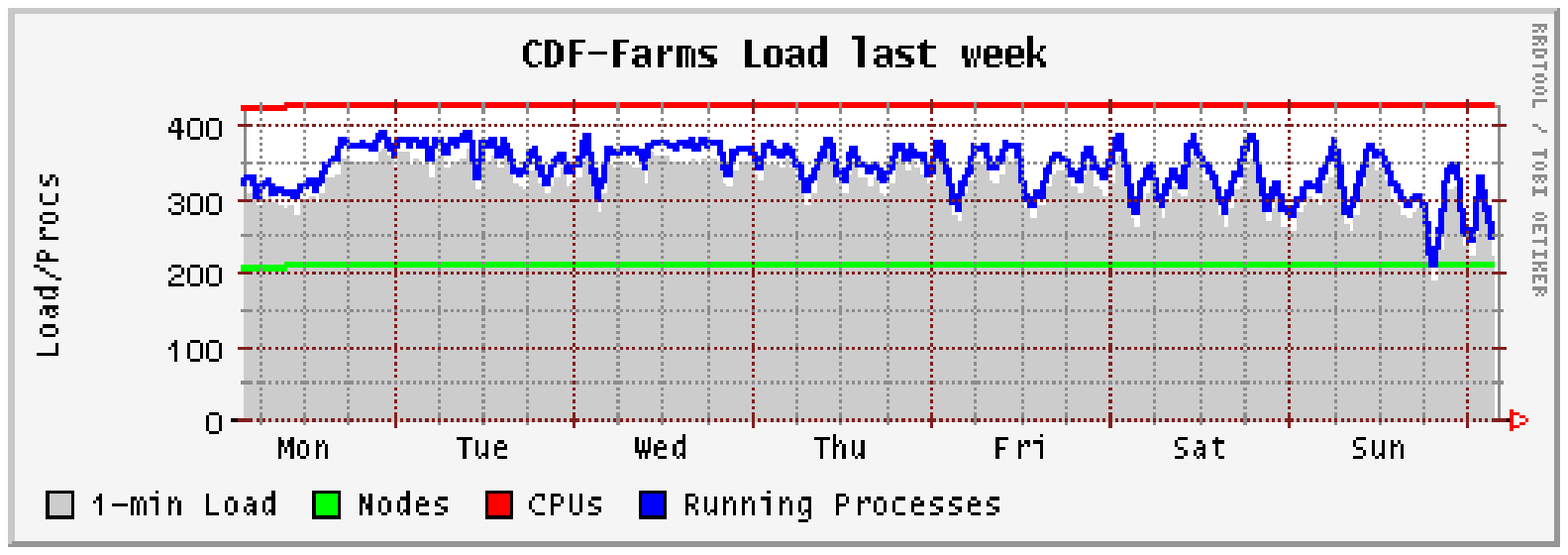,width=1.0\linewidth}
  \caption{SAM farm CPU load (shaded) in production.
  \label{fig:0h_load} }

  \vspace{.5cm}
  \centering\epsfig{file=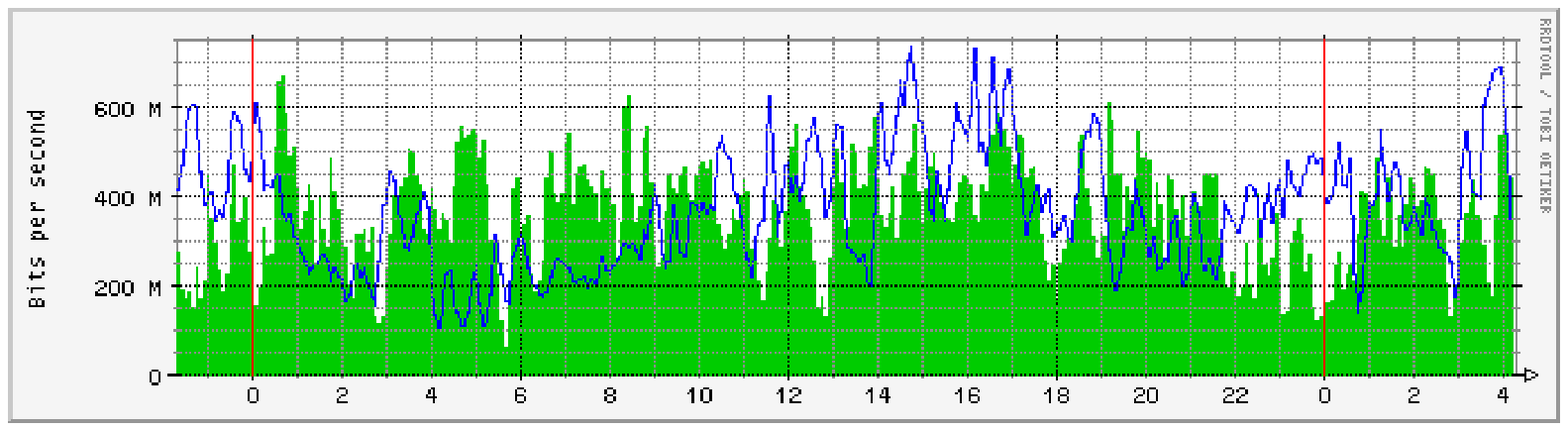,width=1.\linewidth}
  \caption{Network of SAM farm input to durable storage (shaded)
           and output to Enstore (line).
  \label{fig:0h_network} }
\end{figure}

The data through-put was tuned to optimize the use of network, 
durable storage, and the concatenation of two jobs running on the 
durable file servers.
Each data stream in process was limited to about 500 GByte per day.
The total processing speed was kept at full CPU usage,
corresponding to a total of over 20 million events (3 TByte) processed.
Shown in Fig.~\ref{fig:0h_load} is the load of SAM farm CPU 
for a run range of 320 million events.
A typical daily traffic of input and output to Enstore is
shown in Fig.~\ref{fig:0h_network}.
The usage of CPU became less consistent towards the end of the 
production. It was because some small data streams were finished, 
leaving fewer number
of streams in production, and busy traffic writing to 
a smaller number of file servers.

\section{Conclusion}

The CDF production farms have been successfully providing the 
computing processing required for the CDF experiment in Run II.  
The success of this system has in turn enabled successful analysis 
of the wealth of new data being collected by the CDF experiment 
at Fermilab. 
The system has been modified and enhanced during the years of 
its operation to adjust to new requirements and to enable 
new capabilities. It was recently upgraded to adapt to the SAM data 
handling system to be portable on distributed computing facilities.
Its current processing capacity of 20 million events per day
is sufficient for the data taking rate at 40 MByte/sec maximum speed. 
It can be scaled by increasing the number of input data streams
as well as CPU, durable storage, and tape logging.
The production farm system is about a third of the total CDF
computing capacity at Fermilab. 
We may triple the data processing capacity by allocating more 
resources in use.
Also the continuous growth of computing will allow CDF to proceed
in processing and analyzing data through to the end of the life of 
the experiment.

{}

\end{document}